\documentclass[aps,prd,onecolumn,preprintnumbers,groupedaddress,showpacs,nofootinbib,amssymb]{revtex4}
\usepackage[dvipdfm]{graphicx}
\usepackage{bm}
\usepackage{amsmath}
\usepackage{amssymb}
\usepackage{amsfonts}

\begin{document}

\newcommand{\be}{\begin{equation}}
\newcommand{\ee}{\end{equation}}
\newcommand{\bea}{\begin{eqnarray}}
\newcommand{\eea}{\end{eqnarray}}
\newcommand{\beaa}{\begin{eqnarray*}}
\newcommand{\eeaa}{\end{eqnarray*}}
\newcommand{\Lhat}{\widehat{\mathcal{L}}}
\newcommand{\nn}{\nonumber \\}
\newcommand{\e}{\mathrm{e}}
\newcommand{\tr}{\mathrm{tr}\,}

\tolerance=5000

\title{The unification of inflation and late-time acceleration in the frame of $k$-essence}

\author{Rio Saitou$^1$, Shin'ichi Nojiri$^{1,2}$}

\affiliation{
${}^1$Department of Physics, Nagoya University, Nagoya 464-8602, Japan \\
${}^2$Kobayashi-Maskawa Institute for the Origin of Particles and the
Universe, Nagoya University, Nagoya 464-8602, Japan}


\begin{abstract}

By using the formulation of the reconstruction, we explicitly construct models of $k$-essence, 
which unify the inflation in the early universe and the late accelerating expansion 
of the present universe by a single scalar field. Due to the higher derivative terms, 
the solution describing the unification can be stable in the space of solutions, which makes 
the restriction for the initial condition relaxed. 
The higher derivative terms also eliminate tachyon. 
Therefore we can construct a model describing the time development, which cannot be 
realized by a usual inflaton or quintessence models of the canonical scalar field  
due to the instability or the existence of tachyon.  
We also propose a mechanism of the reheating by the quantum effects coming from the variation 
of the energy density of the scalar field. 

\end{abstract}


\maketitle

\section{Introduction \label{SecI}}

We now believe the accelerating expansion of the present universe by several 
cosmological observations \cite{WMAP1,Komatsu:2008hk,SN1,Astier:2005qq}. 
The acceleration has been often supposed to be generated by the dark energy, 
which is an unknown fluid.  
So-called $k$-essence model \cite{Chiba:1999ka,ArmendarizPicon:2000dh,ArmendarizPicon:2000ah} 
is a model of the dark energy. The $k$-essence model is originated 
from $k$-inflation model \cite{ArmendarizPicon:1999rj,Garriga:1999vw}. 
It is possible to regard the tachyon dark energy 
model \cite{Sen:2002nu,Sen:2002an,Gibbons:2002md,Bagla:2002yn}, 
ghost condensation model \cite{ArkaniHamed:2003uy,ArkaniHamed:2003uz}, 
and scalar field quintessence model \cite{Peebles:1987ek,Ratra:1987rm,Chiba:1997ej,Zlatev:1998tr}
as variations of the $k$-essence model. 

Since the $k$-essence model is originated from $k$-inflation model, it might be natural to consider a model 
unifying the inflation and the late acceleration by a single scalar field. 
In this paper, we try to construct such models by using the formulation of the reconstruction 
\cite{Nojiri:2006be,Elizalde:2008yf,Nojiri:2009fq,Nojiri:2009kx,Matsumoto:2010uv,Nojiri:2010wj} and 
we also propose a mechanism of the reheating by the quantum effects. 
In the models, the solution which describes the unification of the inflation and the late acceleration 
can be stable in the space of solutions and also there does not 
appear tachyon due to the higher derivative terms. This tells that we can construct a model 
describing the time development, which cannot be 
realized by models of usual canonical scalar field like inflaton or quintessence 
due to the instability or the existence of tachyon.  

\section{Review of the reconstruction and the stability of the solution \label{SecII}}

In this section, based on \cite{Nojiri:2010wj}, we review on the reconstruction by using  
e-folding $N$, which will be defined in this section, and discuss the stability of the solution in the space of solutions. 
A formulation of the reconstruction using the cosmological time has been given in 
\cite{Matsumoto:2010uv} (about the reconstruction of the canonical/phantom scalar field, 
see \cite{Nojiri:2005pu,Capozziello:2005tf} and about 
the general formalism of the reconstruction, see 
\cite{Nojiri:2006be,Elizalde:2008yf,Nojiri:2009fq,Nojiri:2009kx,Matsumoto:2010uv,Nojiri:2010wj}). 
In the formulation using the cosmological time \cite{Matsumoto:2010uv}, 
it is troublesome and difficult to discuss about the stability of the solution when matters are included. 
In the formulation using the e-folding $N$, as long as the $N$-dependence of the matters 
are known, which is often true as we will see, it is easy to construct a model where 
the solution is stable. 

We now consider a rather general model, whose action is given by
\be
\label{KK1}
S=  \int d^4 x \sqrt{-g} \left( \frac{R}{2\kappa^2} - K \left( 
\phi, X \right) + L_\mathrm{matter}\right)\, ,\quad X \equiv \partial^\mu \phi \partial_\mu \phi \, .
\ee
Here $\phi$ is a scalar field. 
Now the Einstein equation has the following form:
\be
\label{Sch2}
\frac{1}{\kappa^2}\left( R_{\mu\nu} - \frac{1}{2}g_{\mu\nu} R \right) 
= - K \left( \phi, X \right) g_{\mu\nu} 
+ 2 K_X \left( \phi, X \right) \partial_\mu \phi \partial_\nu \phi 
+ T_{\mu\nu}\, .
\ee
Here $K_X \left( \phi, X \right) \equiv \partial K \left( \phi, X \right) / \partial X$ and 
$T_{\mu\nu}$ is the energy-momentum tensor of the matters. 
On the other hand, the variation of $\phi$ gives
\be
\label{Sch3}
0= - K_\phi \left( \phi, X \right) 
+ 2 \nabla^\mu \left( K_X \left( \phi, X \right) \partial_\mu \phi \right)\, .
\ee
Here $K_\phi \left( \phi, X \right) \equiv \partial K \left( \phi, X \right) / \partial \phi$ and 
we have assumed that the scalar field $\phi$ does not directly couple with the matter. 

We now assume the FRW universe whose spacial part is flat: 
\be
\label{FRW}
ds^2 = -dt^2 + a(t)^2 \sum _{i=1,2,3} (dx^i)^2 \, ,
\ee
and the scalar field $\phi$ only depends on time. 
Then the FRW equations are given by 
\be
\label{KK2}
\frac{3}{\kappa^2} H^2 = 2 X \frac{\partial K\left( \phi, X \right)}{\partial X} 
 - K\left( \phi, X \right) + \rho_\mathrm{matter}(t)\, ,\quad 
 - \frac{1}{\kappa^2}\left(2 \dot H + 3 H^2 \right) 
= K\left( \phi, X \right) + p_\mathrm{matter}(t)\, .
\ee
It is often convenient to use redshift $z$ instead of cosmological time $t$ 
since the redshift has direct relation with observations (see \cite{Nojiri:2009kx} for the reconstruction 
of $F(R)$ gravity using the redshift $z$). The redshift is defined by 
\be
\label{redshift}
a(t) = \frac{a\left(t_0\right)}{\left(1+z\right)} = \e^{N - N_0}\, . 
\ee
Here $t_0$ is the cosmological time of the present universe, $N_0$ could be an arbitrary constant, 
and $N$ is called as e-folding and directly related with the redshift $z$. 
In terms of $N$, the FRW equations (\ref{KK2}) can be rewritten as
\be
\label{KK2N}
\frac{3}{\kappa^2} H^2 = 2 X \frac{\partial K\left( \phi, X \right)}{\partial X} 
 - K\left( \phi, X \right) + \rho_\mathrm{matter}(N)\, ,\quad 
 - \frac{1}{\kappa^2}\left(2 H H' + 3 H^2 \right) 
= K\left( \phi, X \right) + p_\mathrm{matter}(N)\, .
\ee
Here $H' \equiv dH/dN$. 
If the matters have constant EoS parameters $w_i$, 
the energy density of the matters is given by 
\bea
\label{KGC1}
&& \rho_\mathrm{matter}(N) = \sum _i \rho_{0i}a^{-3(1+w_i)} 
= \sum_i \rho_{0i}\e^{-3(1+w_i)\left(N-N_0\right)} \, ,\nn
&& p_\mathrm{matter}(N) = \sum _i w_i \rho_{0i}a^{-3(1+w_i)} 
= \sum _i w_i \rho_{0i}\e^{-3(1+w_i)\left(N-N_0\right)} \, .
\eea 
Here $\rho_{0i}$'s are constants. 
Eq.~(\ref{KGC1}) tells the $N$ dependence of the matter energy density 
$\rho_\mathrm{matter}$ is explicitly given. 
Note that the $N$ dependence is not so clear when the matters are created or 
annihilated as in the period of the reheating but in the periods of the inflation 
and the late acceleration, the expression of $\rho_\mathrm{matter}$ in (\ref{KGC1}) could be valid.  
For the general energy density of matters $\rho_\mathrm{matter}(N)$, since the conservation law
\be
\label{KGC2}
\dot \rho_\mathrm{matter} + 3 H \left( \rho_\mathrm{matter} + 
p_\mathrm{matter} \right) = 0\, ,
\ee
can be rewritten in terms of $N$ as 
\be
\label{KGC3}
\rho_\mathrm{matter}'(N)  + 3 \left( \rho_\mathrm{matter}(N) + 
p_\mathrm{matter} (N) \right) = 0\, ,
\ee
we find 
\be
\label{KGC4}
p_\mathrm{matter} (N) = - \rho_\mathrm{matter}(N) - \frac{1}{3} \rho_\mathrm{matter}'(N)\, .
\ee
Then we can rewrite the FRW equations (\ref{KK2N}) as 
\be
\label{KGC5}
K\left( \phi, X \right) = - \frac{1}{\kappa^2}\left(2 H \frac{d H}{dN} + 3 H^2 \right) 
+ \rho_\mathrm{matter}(N) + \frac{1}{3} \rho_\mathrm{matter}'(N) \, ,\quad 
 -X \frac{\partial K\left( \phi, X \right)}{\partial X} 
=  \frac{1}{\kappa^2} H \frac{d H}{dN} - \frac{1}{6} \rho_\mathrm{matter}'(N)\, .
\ee
If we define a new variable $G(N) = H(N)^2$, the equations in (\ref{KGC5}) have the following forms:
\be
\label{KGC6}
K\left( \phi, X \right) = - \frac{1}{\kappa^2}\left( G'(N) + 3 G(N) \right) 
+ \rho_\mathrm{matter}(N) + \frac{1}{3} \rho_\mathrm{matter}'(N) \, ,\quad 
 -X \frac{\partial K\left( \phi, X \right)}{\partial X}
=  \frac{1}{2 \kappa^2} G'(N) - \frac{1}{6} \rho_\mathrm{matter}'(N)\, .
\ee
Then by using the appropriate function $g_\phi (\phi)$, if we choose 
\bea
\label{KGC7}
&& K(\phi,X) = \sum_{n=0}^\infty \left(\frac{X}{g_\phi (\phi) 
+ \frac{\kappa^2}{3} \rho_\mathrm{matter}(\phi)} 
+ 1\right)^n \tilde K^{(n)} (\phi) \, , \nn
&& \tilde K^{(0)} (\phi) \equiv - \frac{1}{\kappa^2}\left( g_\phi'(\phi)  
+ 3 g_\phi(\phi) \right)\, , \quad 
\tilde K^{(1)} (\phi) = \frac{1}{2 \kappa^2} g_\phi'(\phi) \, ,
\eea
we find the following solution for the FRW equations (\ref{KK2}), 
\be
\label{KGC8}
G(N) = H(N)^2 = g_\phi (N) + \frac{\kappa^2}{3} \rho_\mathrm{matter}(N)\, , 
\quad \phi = N \quad \left(X=-H^2\right)\, .
\ee
Now $\tilde K^{(n)} (\phi)$ with $n\geq 2$ can be arbitrary. 
As we will see, $\tilde K^{(2)} (\phi)$ is related with the stability of the 
solution and the existence of tachyon although $\tilde K^{(n)} (\phi)$ with $n\geq 2$ does not 
affect the development of the expansion of the universe. 

We should note that the solution (\ref{KGC8}) is merely one of solutions of the FRW equations 
(\ref{KGC5}) in the model given by (\ref{KGC7}). 
In order that the solution (\ref{KGC8}) could be surely realized, the solution (\ref{KGC8}) 
should be stable under the perturbation in the space of solutions of the FRW equations. 
We now write the perturbation from the solution (\ref{KGC8}) as follows,
\be
\label{KGC9}
G(N) = G_0(N) + \delta G(N) \quad 
\left( G_0(N) \equiv g_\phi (N) + \frac{\kappa^2}{3} \rho_\mathrm{matter}(N) \right) \, , 
\quad \phi = N + \delta \phi(N) \, .
\ee
We should note that in many cases, the $N$-dependence in the energy density 
$\rho_\mathrm{matter}$ of matter is usually given by a fixed function as in (\ref{KGC1}) 
and therefore we find $\delta \rho_\mathrm{matter}=0$. 
Then the equations in (\ref{KGC6}) gives,
\bea
\label{KGC10}
&& - \frac{1}{\kappa^2} \left( g_\phi''(N) + 3 g_\phi'(N) \right) \delta\phi(N) 
 - \frac{g_\phi'(N)}{2\kappa^2} \left(\frac{\delta G(N)}{G_0(N)} + 2\delta \phi'(N) 
 - \frac{G_0'(N)}{G_0(N)} \delta\phi(N) \right) 
= - \frac{1}{\kappa^2} \left( \delta G'(N) + 3 \delta G(N) \right) \, , \nn
&& \frac{1}{\kappa^2} g_\phi'(N) \delta \phi'(N) + \frac{g_\phi'(N)}{2\kappa^2} \frac{\delta G(N)}{G_0(N)} 
+ \frac{g_\phi''(N)}{2\kappa^2} \delta\phi(N) 
 - \frac{g_\phi'(N)}{2\kappa^2} \frac{G_0'(N)}{G_0(N)}\delta\phi(N) \nn
&&  - 2 \tilde K^{(2)} (N) \left(\frac{\delta G(N)}{G_0(N)} + 2\delta \phi'(N)  
 - \frac{G_0'(N)}{G_0(N)} \delta\phi(N) \right) = \frac{1}{2\kappa^2} \delta G'(N)\, .
\eea
Then we find
\bea
\label{KGC11}
&& \left( \begin{array}{c} 
\delta \phi'(N) \\ \delta G'(N) 
\end{array} \right) 
= \frac{1}{L(N)} \left( \begin{array}{cc} A & B \\
C & D \end{array} \right) 
\left( \begin{array}{c} 
\delta \phi(N) \\ \delta G(N) 
\end{array} \right) \, , \nn
&& A \equiv 3 g_\phi'(N) + \frac{G_0'(N)}{2G_0(N)} L(N) \, , \quad 
B \equiv -3 - \frac{L(N)}{2G_0(N)} \, , \nn
&& C \equiv \left(g_\phi''(N) + 3g_\phi'(N) \right) L(N) 
 + 3 g_\phi'(N)^2 \, , \quad 
D \equiv -3L(N) - 3 g_\phi'(N) \, .
\eea
Here
\be
\label{KGC12}
L(N) \equiv g_\phi'(N) - 8\kappa^2 \tilde K^{(2)} (N)\, .
\ee
In order for the solution (\ref{KGC8}) to be stable, the perturbations $\delta \phi(N)$ 
and $\delta G(N)$ should decrease with the increase of $N$, which requires that the real parts of 
the eigenvalues for the matrix $\left( \begin{array}{cc} A & B \\
C & D \end{array} \right)$ should be negative. 
Therefore the stability of the solution requires $A+D < 0$ and $AD - BC > 0$, which gives
\bea
\label{KGC13}
&& 3 > \frac{G_0'(N)}{2G_0(N)} \, , \\
\label{KGC13B}
&& \left( 2G_0(N) + 3L(N) \right) g_\phi''(N) 
+ \left( g_\phi'(N) - G_0'(N) + L(N) \right) g_\phi'(N) - L(N)G_0'(N) > 0 \, .
\eea
We can find $\dot H < 3H^2$ from (\ref{KGC13}), which is always satisfied when the universe 
is in the non-phantom phase, where $\dot H\leq 0$. 
For later convenience, we rewrite (\ref{KGC13B}) in the following form:
\bea
\label{KGC13C}
&& H^2 \tilde K^{(1)\prime} (N) - 2 H H' \tilde K^{(1)}(N) 
+ 3 \kappa^2 \tilde K^{(1)}(N) \tilde K^{(1)\prime}(N) + 2 \left( \tilde K^{(1)}(N) \right)^2 \nn
&& + \left( 4 H H' - 12 \kappa^2 \tilde K^{(1)\prime}(N) - 4 \kappa^2 \tilde K^{(1)}(N) \right) 
\tilde K^{(2)}(N) > 0\, .
\eea
The condition (\ref{KGC13B}) or (\ref{KGC13C}) can be satisfied by choosing $L(N)$ 
and therefore $\tilde K^{(2)} (N)$ properly. 


In case of usual inflaton or quintessence model, where $\tilde K^{(n)} (\phi)=0$ $\left(n\geq 2\right)$ in 
(\ref{KGC7}), there appears tachyon if the potential is concave downwards and therefore the system becomes unstable. 
We now show that the development of the expansion in the universe generated by the concave potential in case 
of the inflaton or quintessence model can be realized without tachyon by adjusting $\tilde K^{(2)} (\phi)$ 
in the $k$-essence models in this paper. 

We now consider the perturbation of only scalar field $\phi$ from the solution (\ref{KGC8}) as 
\be
\label{pp1}
\phi = N + \delta \phi(x^i)\, .
\ee
Different from the case of (\ref{KGC9}), we assume $\delta \phi$ only depends on the spacial coordinate $x^i$ 
since we are now interested in the pole of the scalar field propagator for the spacial momentum, corresponding to tachyon. 
Then by using (\ref{Sch3}) and (\ref{KGC7}), we obtain
\bea
\label{pp2}
0 &=& - 2\frac{\tilde K^{(1)}(N)}{H^2 a^2} \triangle \left(\delta \phi\right) 
+ 2\left\{ \frac{1}{2}{\tilde K^{(0)''}}(N) + \tilde K^{(1)''}(N)
+ \left( 3-\frac{H'}{H}\right){\tilde K^{(1)'}}(N) + \left( -\frac{H''}{H} 
+ \left( \frac{H'}{H} \right)^2  - \frac{6H'}{H} \right) \tilde K^{(1)}(N) \right. \nn
&& \left. + \left( - 4 \left( \frac{H'}{H} \right)^2 + \frac{4H''}{H} + \frac{12 H'}{H} \right) 
\tilde K^{(2)}(N) + \frac{4H'}{H} {\tilde K^{(2)'}}(N) \right\} \delta \phi\, .
\eea
Here $\triangle$ is the Laplacian for the spacial coordinates $x^i$. 
Then if
\bea
\label{pp3}
&& \frac{1}{\tilde K^{(1)}(N)} \left\{ -\frac{H'}{H}\tilde K^{(1)\prime}(N) 
+ \left( -\frac{H''}{H} + \left( \frac{H'}{H} \right)^2 - \frac{6H'}{H} \right) \tilde K^{(1)}(N) \right. \nn
&& \left. + \left( - 4 \left( \frac{H'}{H} \right)^2 + \frac{4 H''}{H} + \frac{12 H'}{H} \right) 
\tilde K^{(2)}(N) + \frac{4 H'}{H} {\tilde K^{(2)\prime}}(N) \right\} \leq 0\, ,
\eea
there does not appear tachyon. 
If we assume $\tilde K^{(1)}(N) < 0$, which corresponds to non-phantom universe, 
and define $\tilde k^{(2)}(N)$ by 
\be
\label{ppk}
\tilde K^{(2)}(N) = H\left( a^3H'\right)^{-1} \tilde k^{(2)}(N)\, ,
\ee
Eq.~(\ref{pp3}) gives
\be
\label{pp4}
\left. \frac{d\tilde k^{(2)}}{d\phi } \right|_{\phi=N} \geq \frac{a^3}{4}\left\{ 
\frac{H'}{H}{\tilde K^{(1)'}}(N) - \left( -\frac{H''}{H} 
+ \left( \frac{H'}{H} \right)^2  - \frac{6H'}{H} \right) \tilde K^{(1)}(N) 
\right\} \, .
\ee
Then if (\ref{KGC13B}) or (\ref{KGC13C}) and (\ref{pp3}) or (\ref{pp4}) 
are satisfied simultaneously without divergence, we obtain a stable model without tachyon.

\section{Models unifying the inflation and the accelerating expansion \label{SecIII}}

In this section, we propose models unifying the inflation in the early universe and the accelerating expansion 
in the present universe. 

\begin{figure}[ht]
\begin{center}

\unitlength=0.6mm
\begin{picture}(120,100)

\thicklines

\put(10,15){\vector(1,0){100}}
\put(10,15){\vector(0,1){70}}

\put(115,15){\makebox(0,0){$N$}}
\put(10,90){\makebox(0,0){$\rho_\phi(N)$}}
\put(7,12){\makebox(0,0){$0$}}

\put(10,77){\line(1,0){45}}
\qbezier(55,77)(63,77)(64,47)
\qbezier(64,47)(65,17)(77,17)
\put(77,17){\line(1,0){33}}

\put(64,10){\makebox(0,0){$N_\mathrm{I}$}}

\thinlines
\put(64,15){\line(0,1){32}}

\end{picture}

\caption{The expected behavior of $\rho_\phi(N)$. \label{Fig1}}
\end{center}
\end{figure}
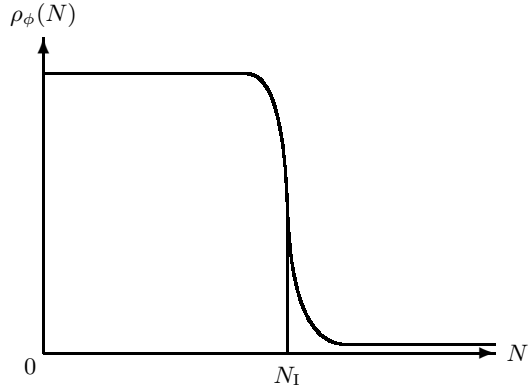

In (\ref{KGC8}), $g_\phi (N)$ corresponds to the energy density $\rho_\phi$ of the scalar field $\phi$:
\be
\label{KN1}
\rho_\phi(N) = \frac{3}{\kappa^2} g_\phi (N) = 2 X \frac{\partial K\left( \phi, X \right)}{\partial X} 
 - K\left( \phi, X \right)\, .
\ee
We expect that the energy density $\rho_\phi$ would behave as the cosmological constant in 
the period of the inflation and the late acceleration. Then we expect the behavior of $\rho_\phi$ 
as in FIG.~\ref{Fig1}. 
We consider the model that the particle production and the reheating would occur after the inflation. 

We now assume 
\begin{enumerate}
\item The energy scale of inflation should be almost equal to the GUT scale.
\item Except the period of the particle production, the EoS parameter $w_\phi$ 
for the scalar field  $\phi$ could be given by
\be
\label{EoSpara}
w_\phi(N) = -1 -\frac{\rho_\phi'(N)}{3\rho_\phi(N)}\, .
\ee
\item In general, there are two solutions $N=N_1$, $N_2$ $\left(N_1<N_2\right)$ in the equation
\be
\label{N12}
\frac{\left. \rho_\phi'\right|_{N= N_{1,2}}}{\left. \rho_\phi'\right|_\mathrm{max}} 
= \frac{1}{\e}\, .
\ee
We expect that the expression (\ref{EoSpara}) could become invalid when $N_1<N<N_2$. 
\item The inflation started at $N=0$ and the end of the inflation is defined by 
$N=N_\mathrm{I}\equiv N_1 \simeq 60$. 
\item The reheating and the particle production could have occurred when $N_1\lesssim N \lesssim N_2$. 
\item The reheating temperature $T_\mathrm{RH}$ could be  
$10\,\mathrm{MeV} < T_\mathrm{RH} < 10^{14}\mathrm{GeV}$. 
\end{enumerate}
Furthermore, the cosmological observations tell, at present, 
1) the energy density of the dark energy is about $10^{-47}\,\mathrm{GeV}^4$, 
2) the temperature of the present universe is $2.725\,\mathrm{K}$ and that at the epoch 
of the decoupling is almost $3000\,\mathrm{K}$ $\left(0.26\,\mathrm{eV}\right)$, which 
give the following constraints
\bea
\label{KN2}
&1.& \rho_\phi (N=0) \simeq 10^{60}\,\mathrm{GeV}^4 \, , \nn
&2.& \rho_\phi (N=N_0) \simeq 10^{-47}\, \mathrm{GeV}^4 \, , \nn
&3.& w_\phi = -1.023 \pm 0.144 \quad \mbox{at $N_0$} \, .
\eea
Here we choose $N_0$ as the e-folding at present universe and the third constraint in (\ref{KN2}) 
comes from SuperNova Legacy Survey (SNLS) date \cite{Astier:2005qq}.

We should note that in the period of the reheating and/or particle production, it is difficult to 
apply the formulation of the reconstruction since the matter energy density is not always given by an 
explicit function of the e-folding $N$. 
In this paper, we approximate the behavior of $\rho_\phi$ in the period of the reheating and/or 
the particle production by the interpolation from the behaviors in the period  
of the inflation and that after the reheating. 

\subsection{Model 1 \label{subsecA}} 
We now consider the following model as model 1:
\be
\label{KN3}
\rho_\phi (N) = M^4 \exp{\left( -\frac{1}{ d^{-1}
+ c_1 \exp{\left(-\frac{N-N_\mathrm{I}}{\Delta_1 }\right)} }\right)} \, ,
\ee
which gives
\be
\label{KN3b}
\rho _\phi '(N) = -\frac{1}{c_1\Delta_1 }
\frac{\exp{\left( -\frac{N-N_\mathrm{I}}{\Delta_1 }\right)} }{\left( 
(c_1d)^{-1} + \exp{\left( -\frac{N-N_\mathrm{I}}{\Delta_1 }\right)} \right) ^2}\rho _\phi(N) \, .
\ee
Here $c_1$, $d$, and $\Delta_1$ are constants and we choose $c_1 \simeq 6.309$ 
and $M\simeq 10^{15}\mathrm{GeV}$. 
Then the assumptions mentioned above and the constraints (\ref{KN2}) give
\bea
\label{KN4}
&1.&\ d \gg 2 \, ,\nn
&2.&\ \exp{\left( -\frac{1}{ d^{-1}+c_1\exp{\left(\frac{N_\mathrm{I}}{\Delta_1 }\right)} }\right)} 
\simeq 1 \, ,\nn
&3.&\ d \simeq 107\ln 10\left[ 1-107 \ln 10\cdot c_1
\exp{\left({-\frac{(N_0-N_\mathrm{I})}{\Delta_1 }}\right)} \right] ^{-1} \, ,\nn
&4.&\ d \leq \frac{1}{c_1}\left( \sqrt{\frac{1}{0.363c_1\Delta_1 }
\exp{\left({-\frac{N_0-N_\mathrm{I}}{\Delta_1 }}\right)} }
 -\exp{\left({-\frac{N_{0}-N_\mathrm{I}}{\Delta_1 }}\right)} \right) ^{-1} \, .
\eea
Since the scale factor $a$ is proportional to the inverse temperature 
$a=\e^{N-N_0}\propto T^{-1}$, we find
\be
\label{KN5}
\e^{N-N_\mathrm{I}} = \frac{a(N)}{a(N_\mathrm{I})} 
\simeq \frac{a(N)}{a(N_\mathrm{RH})} \simeq \frac{T_\mathrm{RH}}{T} \, ,
\ee
and therefore
\be
\label{KN6}
N_0-N_\mathrm{I} \simeq \ln \left( \frac{T_\mathrm{RH}}{3\times 10^{-4}\, \mathrm{eV} }
\right) =24\mbox{--}61 \quad \mathrm{for}\quad T_{\mathrm{RH}} 
= 10\, \mathrm{MeV}\mbox{--}10^{14}\, \mathrm{GeV}\, .
\ee
The second constraint in (\ref{KN4}) tells that the parameter $d$ 
is expressed by the another parameter $\Delta_1$, so in this model 
there remains only one undetermined parameter.
Then in case $T_\mathrm{RH} = 10\,\mathrm{MeV}$, we find $(0,246.4) < (\Delta_1 ,d) < (1.81,247)$ 
and in case $T_\mathrm{RH} = 10^{14}\, \mathrm{GeV}$, $(0,246.4) < (\Delta_1 ,d) < (4.97,248.2)$. 

\begin{figure}[htbp]
\begin{center}
\includegraphics[width=78mm,clip]{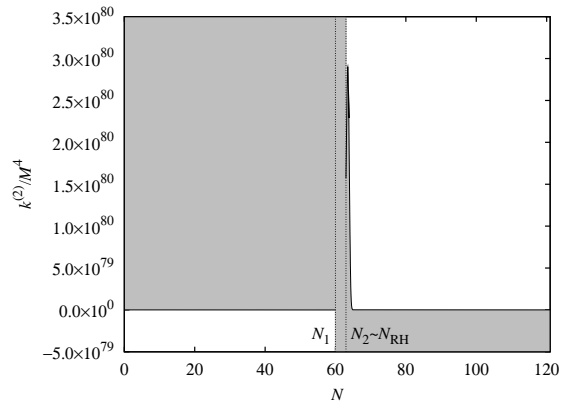}
\end{center}
\caption{The regions satisfying the constraint (\ref{KGC13C}) for $\tilde k^{(2)}(N)$ of model 1. 
The gray regions express the forbidden regions for the instability of the solution. 
In the interval $N_\mathrm{I}=N_1<N<N_2\simeq N_\mathrm{RH}$, the formulation 
of the reconstruction could not be applied due to the particle creation. 
\label{FIG2}}
\end{figure}

\begin{figure}[htbp]
\begin{center}
\includegraphics[width=74mm,clip]{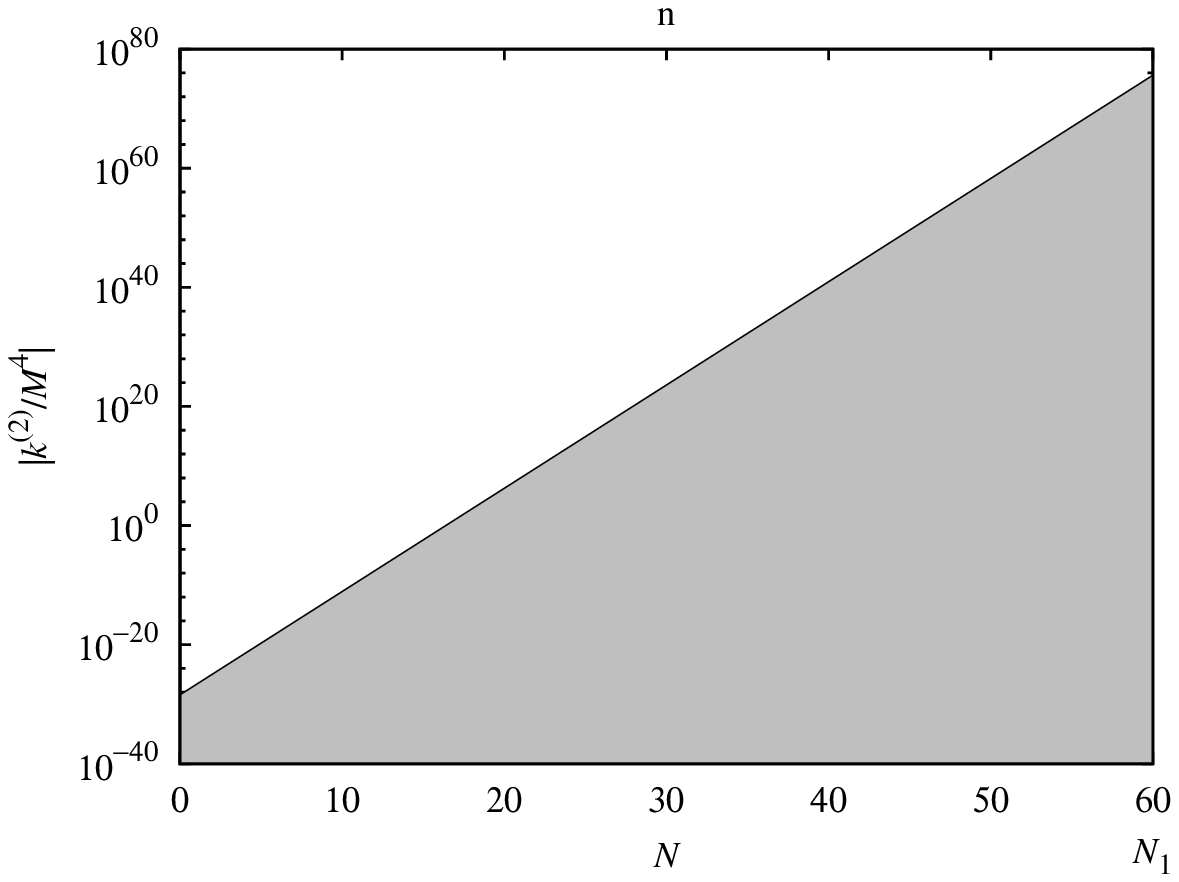}
\end{center}
\caption{The region $0<N<N_\mathrm{I}$ in FIG. \ref{FIG2} is magnified. 
The vertical axis expresses the absolute value of $\tilde k^{(2)}(N)$. 
The symbol `n' means the value of $\tilde k^{(2)}(N)$ is negative there. 
\label{FIG3}}
\end{figure}

\begin{figure}[htbp]
\begin{center}
\includegraphics[width=78mm,clip]{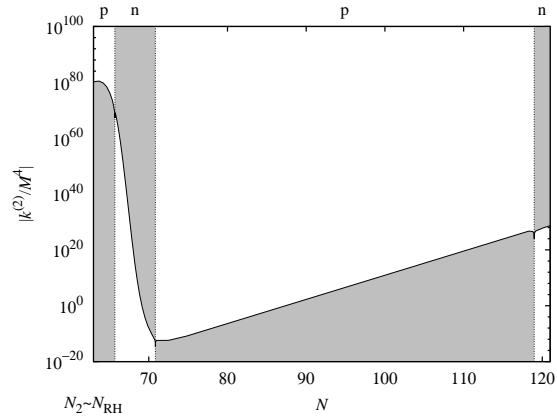}
\end{center}
\caption{The regions $N>N_\mathrm{RH}$ in FIG. \ref{FIG2} are magnified. 
The vertical axis expresses the absolute value 
of $\tilde k^{(2)}(N)$. The symbol `p' means the value of $\tilde k^{(2)}(N)$ 
is positive there. 
\label{FIG4}}
\end{figure}

\begin{figure}[htbp]
\begin{center}
\includegraphics[width=79mm,clip]{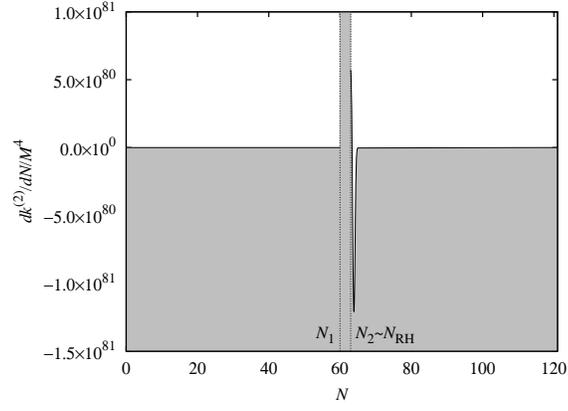}
\end{center}
\caption{The regions satisfying the constraint (\ref{pp4}) for $\tilde k^{(2)\prime}(N)$ of model 1. 
The gray regions express the regions forbidden by the constraint (\ref{pp4}). 
\label{FIG5}}
\end{figure}

\begin{figure}[htbp]
\begin{center}
\includegraphics[width=80mm,clip]{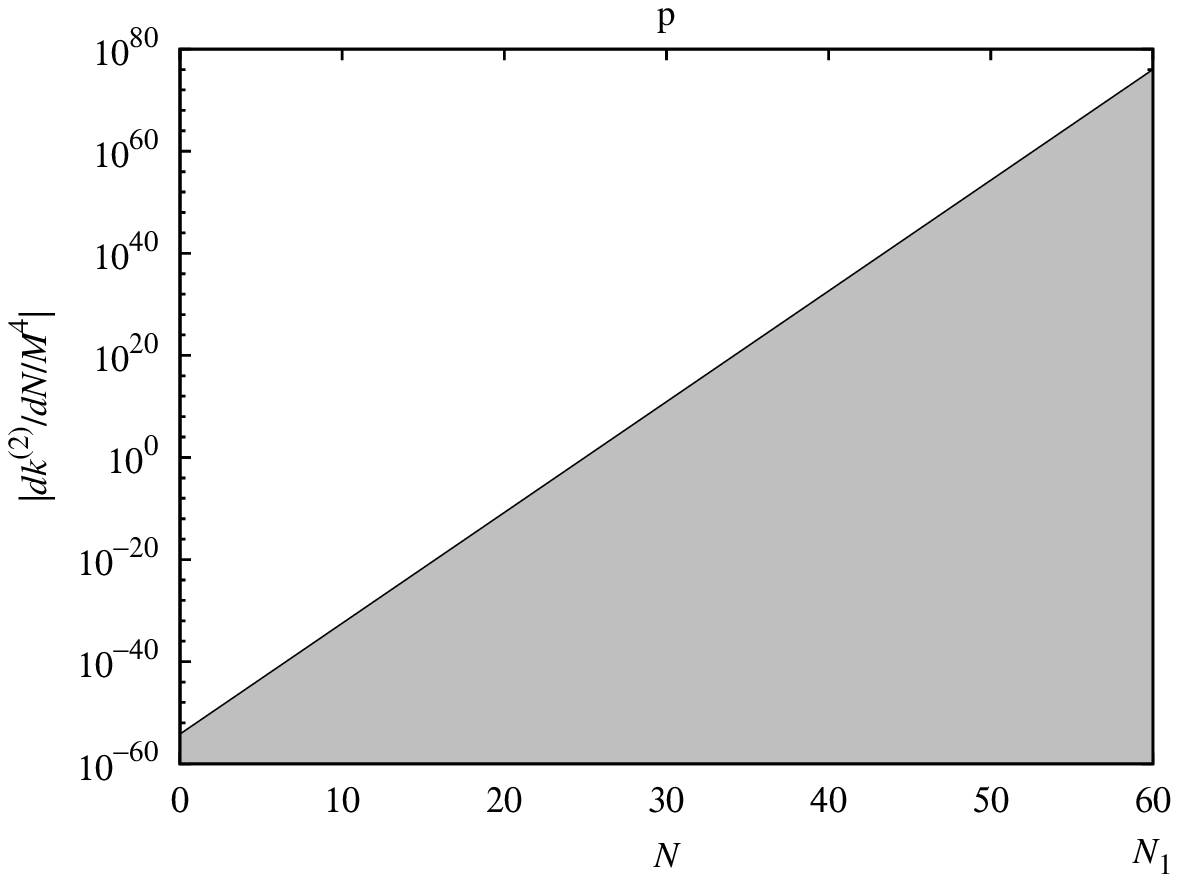}
\end{center}
\caption{The region $0<N<N_\mathrm{I}$ in FIG. \ref{FIG5} is magnified. . 
\label{FIG6}}
\end{figure}

\begin{figure}[htbp]
\begin{center}
\includegraphics[width=80mm,clip]{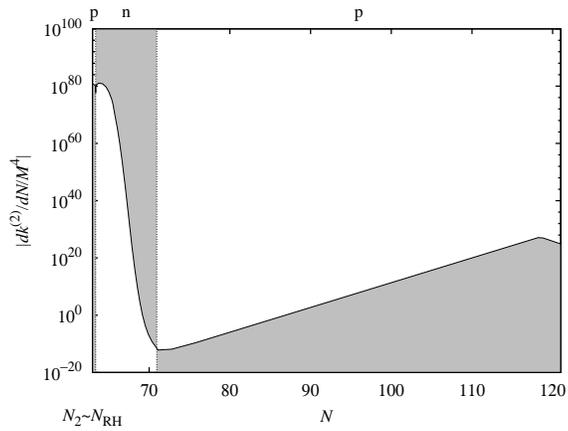}
\end{center}
\caption{The regions $N>N_\mathrm{RH}$ in FIG. \ref{FIG5} are magnified. 
The vertical axis expresses the absolute value of $\tilde k^{(2)\prime}(N)$.  
\label{FIG7}}
\end{figure}

Now the reconstructed action has the following form:
\bea
\label{KN7}
K(\phi, X) &=& \frac{3\tilde K^{(1)}}{\kappa^2\left( \rho_\phi (\phi)+\rho_m(\phi)
\right) }X+\tilde K^{(0)}+\tilde K^{(1)}+\sum_{n=2}^\infty 
\left( \frac{X}{\frac{\kappa ^2}{3}\rho_\phi (\phi )
+\frac{\kappa ^2}{3}\rho _m(\phi )}+1\right) ^n \tilde K^{(n)}(\phi) \, ,\nn
\tilde K^{(0)}(\phi ) &=& -M^4\exp{\left( -\frac{1}{ d^{-1}
+c_1\exp{\left(-\frac{N-N_\mathrm{I}}{\Delta_1 }\right)} }\right)} 
\left( 1-\frac{\exp{\left( -\frac{N-N_\mathrm{I}}{\Delta_1 }\right)} }{3c_1\Delta_1 
\left( (c_1d)^{-1} + \exp{\left( -\frac{N-N_\mathrm{I}}{\Delta_1 }\right)} \right)^2}\right) \, ,\nn
\tilde K^{(1)}(\phi ) &=& -M^4\exp{\left( -\frac{1}{ d^{-1}+c_1
\exp{\left(-\frac{N-N_\mathrm{I}}{\Delta_1 }\right)} }\right)} 
\frac{\exp{\left( -\frac{N-N_\mathrm{I}}{\Delta_1 }\right)} }{6c_1\Delta_1 
\left((c_1d)^{-1} + \exp{\left( -\frac{N-N_\mathrm{I}}{\Delta_1 }\right)} \right)^2} \, .
\eea
In order to find the constraints for $\tilde K^{(2)}(\phi)$ or $\tilde k^{(2)}(\phi )$ given by 
(\ref{KGC13B}) or (\ref{KGC13C}) and (\ref{pp3}) or (\ref{pp4}), we assume 
\begin{align}
&\rho_m(N) \simeq \begin{cases}
0 & \text{for $0\leq N\leq N_\mathrm{I}$} \\
\rho_{m0} \mathrm{e}^{-4(N-N_0)}  & \text{for $N\geq N_\mathrm{RH}\simeq N_2$,}
\end{cases} \nn
&\rho_{m0} \simeq 8.4\times 10^{-52}\mathrm{GeV}^4 \, \ .                 
\end{align} 
Here $N_\mathrm{RH}$ expresses the e-folding number when the reheating finished. 
Then, we obtain approximate constraints for $\tilde k^{(2)}(\phi)$ and $\tilde k^{(2)\prime}(\phi )$ 
for model 1 as shown in FIGs.~\ref{FIG2}--\ref{FIG7} and we can find that there exists 
$\tilde k^{(2)}(\phi)$ or $\tilde K^{(2)}(\phi)$ which satisfies the constraints and does not have 
divergence nor vanish. 

In FIG.~\ref{FIG2}, the region satisfying the constraint (\ref{KGC13C}) is depicted by the directions 
of arrows. When $N_\mathrm{I}=N_1<N<N_2\simeq N_\mathrm{RH}$, 
we could not be able to use the formulation of the reconstruction due to the particle creation. 
The region $0<N<N_\mathrm{I}$ in FIG.~\ref{FIG2} is magnified in FIG.~\ref{FIG3} and the region 
$N>N_2\simeq N_\mathrm{RH}$ in FIG.~\ref{FIG4}. 
The region that $\tilde k^{(2)\prime}(N)$ of model 1 satisfies the constraint (\ref{pp4}) 
is depicted in FIG.~\ref{FIG4} and regions $0<N<N_\mathrm{I}$ and $N>N_2\simeq N_\mathrm{RH}$ 
in FIG.~\ref{FIG4} are magnified in FIG.~\ref{FIG5} and FIG.~\ref{FIG6}, respectively. 
Then we may find that we can always obtain an action where 
the solution becomes stable and does not have tachyon. 

\subsection{Model 2 \label{subsecB}}

As a second model, which we call as model 2, we consider the following:
\be
\label{KN12}
\rho _\phi (N) = \frac{A}{c_2\exp{\left( \frac{N-N_\mathrm{I}}{\Delta_2}\right)} +1}
+B\left( N+b\right)^{-\beta} \, ,
\ee
which gives 
\be
\label{KN12b}
\rho _\phi '(N) = -\frac{A c_2\exp{\left(  \frac{N-N_\mathrm{I}}{\Delta_2}\right)}}{\left(c_2
\exp{\left( \frac{N-N_\mathrm{I}}{\Delta_2}\right)}+1\right)^2}
 -\frac{\beta}{N+b}
B\left( N+b\right)^{-\beta} \, .
\ee
Here $c_2$, $A$, $B$, $\Delta_2$, and $b$ are constants and we choose 
$c_2 \simeq 0.114$ and $A \sim 10^{60}\mathrm{GeV}^4$. 
We now assume that the term with a coefficient $A$ would dominate in the expression of $\rho_\phi$ 
in (\ref{KN2}) when $0<N< N_\mathrm{I}$ and the term with the coefficient $B$ would 
dominate when $N>N_2$. 
Furthermore we choose $A=B$ in order to reduce the number of parameters. 
Then the constraints (\ref{KN2}) give
\bea
\label{KN13}
&1.&\ \frac{1}{8\Delta ^2}\gg \frac{\beta ^2+\beta }{(N_\mathrm{top}+b)^2}(N_\mathrm{top}+b)^{-\beta} \, ,
\quad \frac{c_2}{\Delta _2(c_2+1)^2} \gg \frac{\beta }{N_\mathrm{I}+b}(N_\mathrm{I}+b)^{-\beta} \, ,\nn
&2.&\ 1\gg b^{-\beta} \, ,\nn
&3.&\ \beta \simeq \frac{107\ln 10}{\ln (N_0+b)} \, ,\nn
&4.&\ \beta \leq \left( 0.363 - \frac{109\ln 10-\ln c_2}{100(N_0-N_\mathrm{I})}\right) (N_0+b) \, .
\eea
The second constraint in (\ref{KN13}) tells that the parameter 
$b$ is expressed by another parameter $\beta$, so this model has two undetermined parameters $\beta$ and $\Delta_2$.
Then in case $T_\mathrm{RH} = 10\,\mathrm{MeV}$, we find $(0,0) < (\Delta _2,\beta) < (0.095,47.26)\, (b>99.6)$ 
and in case $T_\mathrm{RH} = 10^{14}\, \mathrm{GeV}$, $(0,0) < (\Delta _2,\beta) < (0.241,49.01)\, (b>31.5)$. 

\begin{figure}[htbp]
\begin{center}
\includegraphics[width=80mm,clip]{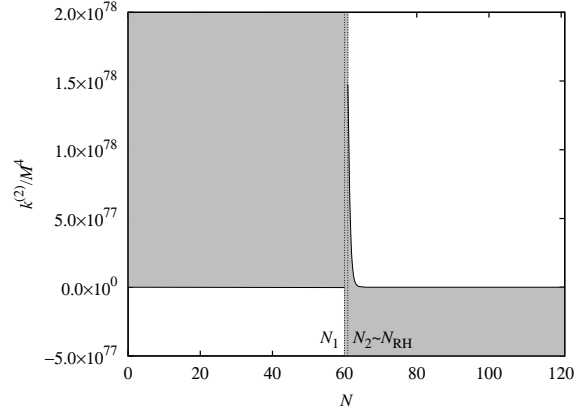}
\end{center}
\caption{The regions satisfying the constraint (\ref{KGC13C}) for 
$\tilde k^{(2)}(N)$ of model 2. 
\label{FIG8}}
\end{figure}

\begin{figure}[htbp]
\begin{center}
\includegraphics[width=80mm,clip]{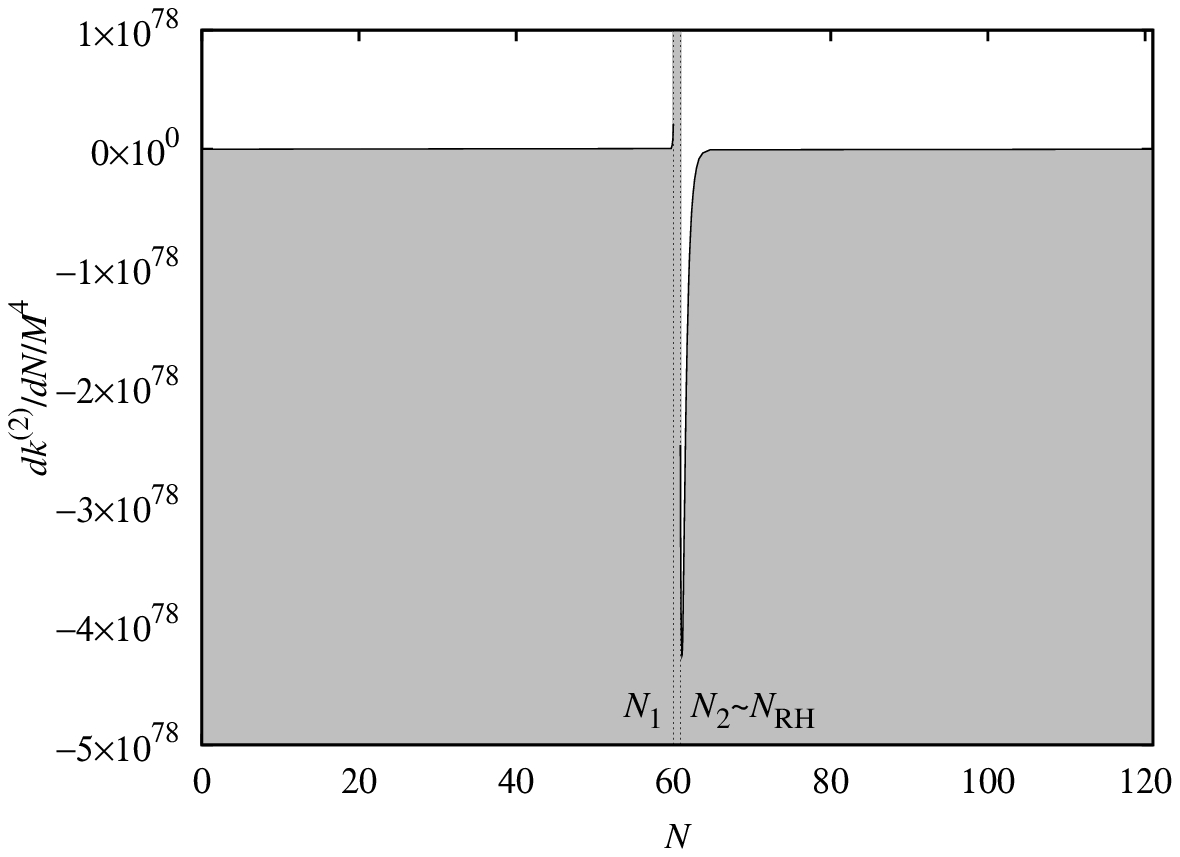}
\end{center}
\caption{The regions satisfying the constraint (\ref{pp4}) for $\tilde k^{(2)\prime}(N)$ of model 2. 
\label{FIG9}}
\end{figure}

Now the reconstructed action has the following form:
\bea
\label{KN14}
K(\phi,X) &=& \frac{3\tilde K^{(1)}}{\kappa^2\left( \rho_\phi (\phi)+\rho_m(\phi)
\right) }X + \tilde K^{(0)} + \tilde K^{(1)} + \sum _{n=2}^\infty 
\left( \frac{X}{\frac{\kappa ^2}{3}\rho_\phi (\phi ) + \frac{\kappa^2}{3}
\rho _m(\phi )} + 1\right) ^n \tilde K^{(n)}(\phi) \, ,\nn
\tilde K^{(0)}(\phi ) &=& -\left( 1-\frac{\frac{c_2}{3}\exp{\left( 
\frac{\phi -N_\mathrm{I}}{\Delta_2}\right)}}{c_2\exp{\left( \frac{\phi-N_\mathrm{I}}{\Delta_2}\right)}
+1}\right) \frac{A}{c_2\exp{\left( \frac{\phi-N_\mathrm{I}}{\Delta_2}\right)} +1 } 
 -\left( 1-\frac{\beta}{3(\phi +b)}A\left( \phi +b\right)^{-\beta}\right) \, ,\nn
\tilde K^{(1)}(\phi ) &=& -\frac{\frac{A c_2}{6}\exp{\left( 
\frac{\phi -N_\mathrm{I}}{\Delta_2}\right)}}{\left(c_2\exp{\left( \frac{\phi -N_\mathrm{I}}{
\Delta_2}\right)}+1\right)^2}
 -\frac{\beta}{6(\phi +b)}A\left( \phi +b\right)^{-\beta} \, .
\eea
Similar to the model 1, by adjusting $\tilde K^{(2)}(\phi)$ of model 2 which does not have divergence 
nor vanish, we obtain a stable model without tachyon. 
In FIG.~\ref{FIG8}, the region satisfying the constraint (\ref{KGC13C}) is depicted 
and the region satisfying the constraint (\ref{pp4}) 
is depicted in FIG.~\ref{FIG9}.

\subsection{The dynamics of the scalar field \label{subsecC}}

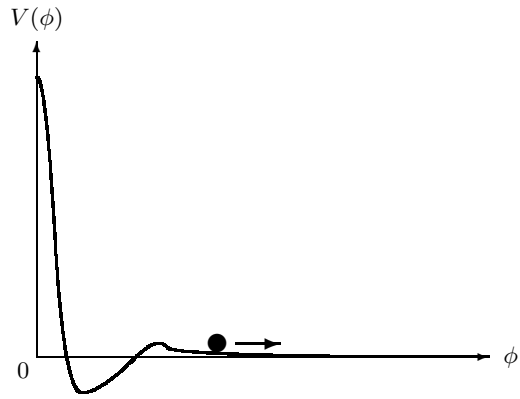
\begin{figure}[htbp]
\begin{center}

\unitlength=0.6mm
\begin{picture}(120,100)

\thinlines

\put(10,15){\vector(1,0){100}}
\put(10,15){\vector(0,1){70}}

\thicklines

\put(115,15){\makebox(0,0){$\phi$}}
\put(10,90){\makebox(0,0){$V(\phi)$}}
\put(7,12){\makebox(0,0){$0$}}

\qbezier(10,77)(12,77)(14,42)
\qbezier(14,42)(16,7)(20,7)
\qbezier(20,7)(24,7)(32,15)
\qbezier(32,15)(35,18)(37,18)
\qbezier(37,18)(38,18)(39,17)
\qbezier(39,17)(41,15)(110,15)

\put(50,18){\circle*{4}}
\put(54,18){\vector(1,0){10}}

\end{picture}

\end{center}
\caption{The effective potential of the canonical scalar field. 
\label{FIG10}}
\end{figure}

The evolution of the expansion in universe does not change even if we consider the model with 
$\tilde K^{(n)}=0$ $(n\geq 2)$, which corresponds to the usual inflaton and/or quintessence models 
since the time evolution of the system is controlled only by $\tilde K^{(0)}(\phi)$ and $\tilde K^{(1)}(\phi)$. 
In case of $\tilde K^{(n)}=0$ $(n\geq 2)$, the scalar field becomes canonical and 
the dynamics of the scalar field is compared with the dynamics of 
a classical particle in a potential. 
In case of $\tilde K^{(n)}=0$ $(n\geq 2)$, in order to generate the development of the universe 
expansion given by in the previous Subsections \ref{subsecA} and \ref{subsecB}, 
the potential has typically the form depicted in FIG.~\ref{FIG10}.

As an initial condition, the scalar field should almost stay near the top of the potential 
in order to generates the inflation. 
After that, it rolls down to the bottom of the potential and creates the particles. 
Finally, without getting trapped in the bottom of the potential, the scalar field goes 
through the subsequent small peak of the potential 
and plays the roll of the dark energy.

It is important that different from the inflaton or quintessence models, 
we need not to fine-tune the initial conditions for the scalar field 
and there are no tachyonic instability in the models 
we have constructed even if the effective potential is concave downwards 
since the motion of the scalar field can be stabilized by their $\tilde K^{(2)}(\phi)$ 
term which should not vanish.

\section{A mechanism of the particle production \label{SecV}}

Now we assume the Hubble rate is given in terms of the e-folding $N$ as
$H=H(N)$ and consider the situation that the e-folding $N$ can be identified with 
a scalar field $\phi$. 
We now consider the interaction between the scalar field $\phi$ between another 
real scalar field $\varphi$ as follows,
\be
\label{p1}
H_\mathrm{int} = -\frac{C_0}{2} \int d^3 x \sqrt{-g} 
\frac{d\rho_\phi(\phi )}{d\phi } \varphi^2\, .
\ee
Here $C_0$ is a constant. 
Note that $\rho_\phi(\phi)$ is not the real energy density of $\phi$ but merely a function 
of $\phi$ given by replacing $N$ in $\rho_\phi (N)$ in (\ref{KN1}) by $\phi$. 
We assume that $\phi$ can be treated as an external source and the interaction 
occurs only in a narrow region around $t=0$ and we approximate 
$C_0 \frac{d\rho_\phi(\phi )}{d\phi }$ as a function of the cosmological time $t$. 
We now approximate $C_0 \frac{d\rho_\phi(\phi )}{d\phi }$ by the Gauss function:
\be
\label{p2}
 - C_0 \frac{d\rho_\phi (\phi )}{d\phi } = \frac{U_0}{\Delta \sqrt{\pi}} \e^{- \frac{t^2}{\Delta^2}}\, .
\ee
Here $U_0$ is a constant and $\Delta$ is the standard deviation. 
We also assume that the space-time can be regarded as static and also flat 
when $\left|t\right| \sim \Delta$, which should be checked. 

Then the amplitude that the vacuum could transit to two-particle state whose momenta 
are given by $\bm{p}$ and $\bm{q}$ is given by
\bea
\label{p3}
A_{\bm{p}\bm{q}} &=& i \int_{-\infty}^\infty dt \left< \bm{p}, \bm{q} \left| 
H_\mathrm{int} \right| 0 \right> \nn
&=& i\frac{U_0}{\Delta \sqrt{\pi}} \int_{-\infty}^\infty dt \int d^3 x 
\frac{ \e^{- \frac{t^2}{\Delta^2} -i \left( \omega_p + \omega_q \right) 
+ i \left(\bm{p} + \bm{q} \right)\cdot \bm{x}}}{2\sqrt{\omega_p \omega_q}} \nn
&=& i \delta^3 \left(\bm{p} + \bm{q} \right) \frac{U_0 \e^{-\Delta^2\omega_p^2}}{2 \omega_p}\, .
\eea
Here $\omega_p = \sqrt{\bm{p}^2 + m_\varphi^2}$ with the mass $m_\varphi$ of $\varphi$. 
Then the transition probability is given by 
\be
\label{p4}
P_2 = \frac{1}{2} \int d^3p\, d^3 q\, \delta^3 \left(\bm{p} + \bm{q} \right)^2 
\frac{U_0^2 \e^{-2 \Delta^2 \omega_p^2}}{4 \omega_p^2}  \nn
= \frac{VU_0^2}{16\pi^2}\int p^2 dp \frac{\e^{-2 \Delta^2 \omega_p^2}}{\omega_p^2} \, .
\ee
The factor $1/2$ in the first line appears since 
$\left< \bm{p}, \bm{q} \right| = \left< \bm{q}, \bm{p} \right|$ and $V$ is the volume of space 
which appears since
\be
\label{p5}
\delta^3(0) = \frac{V}{\left(2\pi\right)^3}\, .
\ee
Especially when $\varphi$ is massless, that is, $m_\varphi=0$, we find
\be
\label{p6}
P_2 = \frac{VU_0^2}{8 \left(2\pi\right)^{\frac{3}{2}} \Delta}\, ,
\ee
which diverges when $\Delta \to 0$. 

Then the total transition probability per unit volume is given by
\be
\label{p7}
p_2 = \frac{P_2}{V}\, .
\ee
Therefore the particle density $n$ is given by
\be
\label{p7b}
n=2p_2 \, .
\ee

We now consider about the energy (density). Eq.~(\ref{p3}) tells that 
the expectation value of the energy $E_2$ corresponding to two particles state is 
given by
\be
\label{p9}
E_2 = \frac{1}{2} \int d^3p\, d^3 q\, \delta^3 \left(\bm{p} + \bm{q} \right)^2 
\frac{2 \omega_p U_0^2 \e^{-2 \Delta^2 \omega_p^2}}{4 \omega_p^2}  \nn
= \frac{VU_0^2}{8\pi^2}\int p^2 dp \frac{\e^{-2 \Delta^2 \omega_p^2}}{\omega_p} \, .
\ee
Especially when $\varphi$ is massless, we find
\be
\label{p10}
E_2 = \frac{VU_0^2}{8 \left(2\pi\right)^{\frac{3}{2}} \Delta}\, .
\ee
Then the expectation value of the energy density $\epsilon_2$ 
for the two particle state is given by 
\be
\label{p11}
\epsilon_2 = \frac{E_2}{V}\, .
\ee

We may estimate the width $\Delta$ in (\ref{p2}) 
by using $N_1$ and $N_2$ in (\ref{N12}) as 
\be
\label{KN8}
\Delta N = N_2-N_1 = \int H\, dt \simeq H_\mathrm{I}\int dt = 2H_\mathrm{I}\Delta \, ,
\ee
which gives $\Delta N \simeq 2.98\Delta_1$ for model 1 and $\Delta N \simeq 4.34\Delta _2$ for model 2. 
Then since the energy density of the radiation in the present universe is given by 
the product of the critical density $\rho_{\mathrm{cr}0}$ and the density parameter $\Omega_{\mathrm{r}0}$ for the radiation. 
Since 
\be
\label{KN9}
\rho_{\mathrm{cr}0} = 10^{-47}\, \mathrm{GeV}^4 \, , \quad \Omega_{\mathrm{r}0} = 8.4\times 10^{-5} \, ,
\ee
we find 
\be
\label{KN10}
\epsilon_2 = \Omega_{\mathrm{r}0}\rho_{\mathrm{cr}0}\left( \frac{T_\mathrm{RH}}{T_0}\right)^4 
= 10.4\times 10^{-10}\mbox{--}10^{54} \, \mathrm{GeV}^4 \simeq \frac{4\times 10^{22}\, 
\mathrm{GeV}^2U_0^2}{8(2\pi )^{3/2}\Delta N^2} \, ,
\ee
which tells 
\be
\label{KN11}
U_0 = 9.86\Delta _1\times 10^{-15}\mbox{--}10^{17}\, \mathrm{GeV} \quad \mbox{for model 1}\, ,\quad 
1.43\Delta _2\times 10^{-14}\mbox{--}10^{18} \, \mathrm{GeV} \quad \mbox{for model 2}\, .
\ee
In (\ref{KN10}), $T_0$ is the temperature of the present universe. 

\section{Summary \label{SecVI}}

In this paper, after reviewing the formulation of reconstruction for $k$-essence, 
we explicitly constructed two models which unify the inflation in the early universe and 
the late-time acceleration in the present universe, and satisfy the observational constraints.
We have proposed a mechanism of the interaction for particle production by the quantum effects 
coming from the variation of the energy density of the scalar field 
and estimated the energy density of the particles. 

In both of the models, the solutions describing the development of the universe expansion 
are stabilized by $\tilde K^{(2)}$ or $\tilde k^{(2)}$ terms which should not vanish. 
We also note that $\tilde K^{(2)}$ or $\tilde k^{(2)}$ terms play the role to eliminate the tachyon. 
As explained in Subsection \ref{subsecC}, the solutions describing the development of the expansion in 
our models can be realized by the usual inflaton or quintessence model, 
where the scalar field is canonical, but in the canonical scalar models, the solutions could be often unstable 
and there could appear a tachyon when the scalar field lies at the concave part of the potential. 
The instability of the canonical scalar models require the fine tuning of the initial conditions, 
which makes the models unnatural. In our models, due to the stability of the solutions controlled by 
the $\tilde K^{(2)}$ or $\tilde k^{(2)}$ terms, there could exist a wide region of the possible initial conditions. 
Then in the framework given in this paper, we can construct a model describing the time development, which cannot be 
realized by a usual inflaton or quintessence model. 

The roles of $\tilde K^{(n)}$ ($n \geq 3$) are, however, still unclear although these terms play the role to 
guarantee the existence of the Schwarzschild solution \cite{Nojiri:2009fq,Nojiri:2010wj}. 
More detailed cosmological constraints may restrict the form of these terms. 
It might be interesting to consider the reconstruction of the general spherical symmetric solution 
in the $k$-essence model as in \cite{Nojiri:2006jy}.

\section*{Acknowledgments}

We are indebted to S.~D.~Odintsov and K.~Bamba for the useful discussion. 
This work is supported in part by Global COE Program
of Nagoya University (G07) provided by the Ministry of Education, Culture, 
Sports, Science \& Technology and 
the JSPS Grant-in-Aid for Scientific Research (S) \# 22224003 (S.N.).

\end{document}